\documentstyle[12pt,bezier]{report}

\makeatletter

\def\refname{References}
\def\thebibliography#1{\chapter*{\refname\@mkboth
  {\uppercase{\refname}}{\uppercase{\refname}}}\list
  {\@biblabel{\arabic{enumiv}}}{\settowidth\labelwidth{\@biblabel{#1}}%
    \leftmargin\labelwidth
    \advance\leftmargin\labelsep
    \itemsep 2pt plus 1pt minus 1pt
    \parsep 0pt plus 2pt
    \usecounter{enumiv}%
    \let\p@enumiv\@empty
    \def\theenumiv{\arabic{enumiv}}}%
    \def\newblock{\hskip .11em plus.33em minus.07em}%
    \sloppy\clubpenalty4000\widowpenalty4000
    \sfcode`\.=1000\relax}

\def\@biblabel#1{#1.\hfill}
\def\@cite#1#2{\unskip$^{\mbox{\scriptsize #1}}$}
\def\@citere#1#2{\unskip\ Ref.~#1}
\def\citere#1{{\let\@cite\@citere\cite{#1}}}
\def\@citeres#1#2{\unskip\ Ref.s~#1}
\def\citeres#1{{\let\@cite\@citeres\cite{#1}}}

\@addtoreset{footnote}{page}  
\def\thefootnote{\fnsymbol{footnote}}

\long\def\@makefntext#1{\parindent 0em\noindent
            \hbox to 0em{\hss$^{\@thefnmark}$}#1}
\long\def\@makefntext#1{\parindent 1em\noindent
            \hbox to 1.8em{\hss$^{\@thefnmark}$}#1}

\let\@eqnsel=\relax

\def\@makechapterhead#1{             
  { \vskip 4.2ex plus 1ex minus .2ex
    \parindent 0pt \raggedright
      \normalsize \bf \thechapter. #1      
    \par
    \nobreak                         
    \vskip 2.0ex plus .2ex           
  } }

\def\@makeschapterhead#1{             
  { \vskip 4.2ex plus 1ex minus .2ex
    \parindent 0pt \raggedright
      \normalsize\bf         #1      
    \par
    \nobreak                         
    \vskip 2.3ex plus .2ex           
  } }

\def\chapter{
   \@afterindenttrue        
   \secdef\@chapter\@schapter}   

\def\@chapter[#1]#2{\ifnum \c@secnumdepth >\m@ne
        \refstepcounter{chapter}
        \typeout{\@chapapp\space\thechapter.}
        \addcontentsline{toc}{chapter}{\protect
        \numberline{\thechapter}#1}\else
      \addcontentsline{toc}{chapter}{#1}\fi
   \chaptermark{#1}
   \addtocontents{lof}{\protect\addvspace{10pt}} 
   \addtocontents{lot}{\protect\addvspace{10pt}} 
   \@makechapterhead{#2}
          \@afterheading               
     }                                 

\def\@schapter#1{\@makeschapterhead{#1}
              \@afterheading}

\@ifundefined{mathrm}{\def\mathrm#1{{\rm #1}}}{\relax}

\def\refeq#1{\mbox{Eq.~(\ref{#1})}}

\long\def\@caption#1[#2]#3{\par\addcontentsline{\csname
  ext@#1\endcsname}{#1}{\protect\numberline{\csname
  the#1\endcsname}{\ignorespaces #2}}\begingroup
    \@parboxrestore
    \footnotesize
    \@makecaption{\csname fnum@#1\endcsname}{\ignorespaces #3}\par
  \endgroup}

\makeatother

\def\beq{\begin{equation}}
\def\eeq{\end{equation}}
\def\beqar{\begin{eqnarray}}
\def\eeqar{\end{eqnarray}}
\def\bma{\begin{displaymath}}
\def\ema{\end{displaymath}}
\def\barr#1{\begin{array}{#1}}
\def\earr{\end{array}}
\def\bit{\begin{itemize}}
\def\eit{\end{itemize}}
\def\bfi{\begin{figure}}
\def\efi{\end{figure}}
\def\bce{\begin{center}}
\def\ece{\end{center}}

\def\si{\sigma}
\def\la{\lambda}

\def\ga{\gamma}

\def\reffi#1{\mbox{Fig.~\ref{#1}}}

\newcommand{\GeV}{\unskip\,\mathrm{GeV}}
\newcommand{\TeV}{\unskip\,\mathrm{TeV}}

\def\mathswitchr#1{\relax\ifmmode{\mathrm{#1}}\else$\mathrm{#1}$\fi}
\newcommand{\PW}{\mathswitchr W}
\newcommand{\PZ}{\mathswitchr Z}

\newcommand{\PH}{\mathswitchr H}
\newcommand{\Pe}{{\mathswitchr e}}
\newcommand{\Pmy}{{\mathswitch \mu}}
\newcommand{\Pta}{{\mathswitch \tau}}

\newcommand{\Pd}{\mathswitchr d}
\newcommand{\Pu}{\mathswitchr u}
\newcommand{\Ps}{\mathswitchr s}
\newcommand{\Pc}{\mathswitchr c}
\newcommand{\Pb}{\mathswitchr b}
\newcommand{\Pt}{\mathswitchr t}
\newcommand{\Pep}{\mathswitchr {e^+}}
\newcommand{\Pem}{\mathswitchr {e^-}}
\newcommand{\PWp}{\mathswitchr {W^+}}
\newcommand{\PWm}{\mathswitchr {W^-}}

\def\mathswitch#1{\relax\ifmmode#1\else$#1$\fi}
\newcommand{\MW}{\mathswitch {M_\PW}}

\newcommand{\MH}{\mathswitch {M_\PH}}

\newcommand{\Mt}{\mathswitch {m_\Pt}}

\newcommand{\Oa}{\mathswitch{{\cal{O}}(\alpha)}}
\newcommand{\Oaa}{\mathswitch{{\cal{O}}(\alpha^2)}}
\newcommand{\M}{{\cal{M}}}

\def\eeffff{\Pep\Pem\to4f}
\def\GW{\Gamma_\PW}

\newsavebox{\wigr}
\newsavebox{\wigu}
\newsavebox{\wigur}
\newsavebox{\Vu}
\newsavebox{\Vr}
\newsavebox{\Fr}
\newsavebox{\Fl}
\newsavebox{\Fu}
\newsavebox{\Vur}
\newsavebox{\Fur}
\newsavebox{\Fdr}
\newsavebox{\Lur}
\newsavebox{\Ldr}

\begin{document}
\thispagestyle{empty}
\def\thefootnote{\fnsymbol{footnote}}
\setcounter{footnote}{1}
\null
\begin{flushright}
 CERN-TH.6928/93 \\
 DESY-93-079
\end{flushright}
\vskip 1cm
\vfil
\begin{center}
{\Large \bf Theoretical Predictions for W-Pair Production%
\footnote{Talk given at the Workshop on Physics and Experiments
with Linear \Pep\Pem~colliders, Waikoloa, Hawaii, April 26--30, 1993.}
\par} \vskip 2.5em
{\large
{\sc W.\ Beenakker} \\[1ex]
{\it Theory Division, DESY, Hamburg, Germany} \\[2ex]
{\sc A.\ Denner}  \\[1ex]
{\it Theory Division, CERN, Geneva, Switzerland}
\par} \vskip 1em
\end{center} \par
\vskip 4cm
\vfil
{\bf Abstract} \par
We review the status of theoretical predictions for \PW-pair production
at high energies. We discuss a systematic scenario towards a Monte-Carlo
generator for $\eeffff(\ga)$, which meets the experimental requirements.
In particular we summarize the recent developments in this field.
\par
\vskip 2cm
\noindent CERN-TH.6928/93 \par
\vskip .15mm
\noindent June 1993 \par
\null
\setcounter{page}{0}
\clearpage

\newfont{\footfont}{cmr9}
\newcommand{\sfootnote}[1]{\footnote{\footfont #1}}

\vspace*{9mm}
\bce
{
\bf THEORETICAL PREDICTIONS FOR W-PAIR PRODUCTION
} \ece
\vspace{6mm}
\bce
\normalsize W.~BEENAKKER$^{\dag}$ and  A.~DENNER$^{\ddag}$ \\[0ex]
{\small {\it $^{\dag}$ DESY, Theory Division, Notkestrasse 85,
D-22603 Hamburg, Germany}}\\
{\small {\it $^{\ddag}$ CERN, Theory Division, CH-1211 Geneva 23,
Switzerland}}
\ece
\vspace{.1ex}
\vspace{-4mm}

\chapter{Introduction}

The gauge-boson production processes allow an accurate direct study
of triple and quartic gauge-boson couplings. Owing to
the presence of unitarity cancellations for longitudinal gauge bosons
at high energies, the sensitivity to anomalous gauge couplings, which in
general spoil these cancellations, grows with energy.

For $\Pep\Pem\to\PWp\PWm$, the most prominent gauge-boson production
process, the sensitivity to anomalous couplings is roughly given by
$\beta^2t^2/s^2\times s/\MW^2$, with $\beta=\sqrt{1-4\MW^2/s}$.
Here the first factor originates from the suppression of the
$s$-channel diagrams, containing the triple gauge couplings, compared to
the dominant $t$-channel diagram. The second factor is due to the
enhancement in the absence of unitarity cancellations. Accordingly
one should go to energies as high as possible%
\sfootnote{For LEP200 the highest energy possible should be aimed at
in order to overcome the $\beta^2$ suppression.}
and consider observables that are not dominated by the $t$-channel pole.
In view of the latter, one needs to consider angular distributions
of the \PW~bosons and their decay products,
which in particular allow to derive information on the polarization
of the \PW~bosons.

As the \PW~bosons decay mainly into fermion--antifermion pairs,
one has to consider the process $\eeffff(\ga)$.
For this one needs a (fast) Monte-Carlo generator
which includes anomalous couplings and has
an accuracy of better than 1\%,
in order to allow theoretical predictions with an uncertainty
\looseness 1
below the experimental precision.

The lowest-order process $\eeffff$ involves diagrams with internal
\PW- and \PZ-boson propagators, which may become resonant and yield
the dominant contributions.
In order to define gauge-invariant resonant contributions and to
introduce finite width effects in a gauge-invariant way,
we adopt the pole scheme \cite{pole scheme}.
This is based on a split-up of the matrix elements according to
the poles of the \PW- and \PZ-boson propagators with corresponding
constant residues \cite{Mor93}. As there can be two resonant
gauge bosons in $\eeffff$, there are double-pole,
single-pole, and non-resonant contributions.
The dominant contributions are given by the double-pole terms.%
\sfootnote{%
It should be noted that below the \PW-pair production
threshold the doubly-resonant contributions vanish and the
singly-resonant ones become dominant.
Moreover there are problems when defining the pole-scheme split-up
in the vicinity of the threshold \cite{Ae93}.
This is, however, not relevant at high energies.}
Since the \PZ-boson double-pole terms are suppressed by roughly a factor
of 10 relative to the \PW-boson ones,
we focus on the latter. In that case
the doubly-resonant diagrams can be related to on-shell \PW-pair
production in the following way:
\beq \label{Mfactor}
\M_{\eeffff}^{\mathrm{pole}} = \sum_{\la_+,\la_-}
\M_{\Pep\Pem\to\PWp\PWm}^{\la_+,\la_-}
\times\frac{\M_{\PWp\to f_1\bar f_2}^{\la_+}}{k_+^2-\MW^2+i\MW\GW}
\times\frac{\M_{\PWm\to f_3\bar f_4}^{\la_-}}{k_-^2-\MW^2+i\MW\GW},
\eeq
where $\M_{\Pep\Pem\to\PWp\PWm}$, $\M_{\PWp\to f_1\bar f_2}$,
$\M_{\PWm\to f_3\bar f_4}$ denote the matrix elements for the
production of two on-shell \PW~bosons and their subsequent
decay into fermion--antifermion
pairs, and $\la_\pm$ the helicities of the \PW~bosons.
The finite width effects are included by introducing the physical width
$\GW$ into the resonant propagators.

At the cross-section level, all other (non-doubly-resonant)
contributions are typically suppressed by a factor $\GW/\MW\approx2.5\%$
for each non-re\-so\-nant \PW~pro\-pa\-ga\-tor \cite{Ae93,Ae91}.
The non-resonant contributions can be further reduced by a
cut on the invariant masses of the decay products
$\MW -\Delta < \sqrt{k_\pm^2} < \MW + \Delta$,
which typically suppresses a flat
background by an additional factor $\Delta/\MW$.
For small enough $\Delta$ this also applies to the background from
\PZ-pair production contributions.

\chapter{Radiative Corrections}

As far as the \Oa~corrections are concerned, it is most likely
sufficient to take into account only the doubly-resonant contributions,
as the others are suppressed by an additional factor $\GW/\MW$.
The only exceptions might be enhanced corrections, which can usually
be treated by renormalization group methods (leading collinear QED
logarithms, running $\alpha$, \ldots) and which can thus simply be
combined with the lowest-order cross-section.

There are two different sources of double-pole terms in the virtual
corrections, factorizable corrections and non-factorizable photonic
corrections.
The factorizable corrections consist of all those diagrams where
the \PW-production and decay parts can be separated by cutting the
two resonant \PW~propagators, i.e.\ the ones that are reducible at both
\PW~lines. Consequently they can be related to the on-shell
matrix elements according to \refeq{Mfactor}. At \Oa\
they comprise the complete resonant non-photonic corrections and the
most important resonant photonic corrections, in particular those
involving leading logarithms.
The on-shell matrix elements are known, including the complete
\Oa~corrections \cite{onshell}.
Thus regarding the factorizable corrections the only task left is their
implementation into a Monte-Carlo generator. For the resonant
diagrams contributing to the \Oa~real photonic corrections, such a
generator already exists \cite{Ae91}.

Feynman diagrams that are not reducible at both \PW~lines
do in general not yield doubly-resonant contributions.
This does not hold for non-factorizable contributions resulting
from diagrams where a virtual photon is exchanged between the two
decay parts or between the production and a decay part of the
diagram (see \reffi{FIphodia}).
These diagrams contain resonant contributions related to the
IR limit, i.e.\ to those parts of the integration region where
both the energy and the momentum of the massless photon are zero.
\bfi
\unitlength 1pt
\savebox{\wigur}(12,7)[bl]
   {\bezier {20}(0.0,0.0)(1.5,5.5)(6.0,3.5)
    \bezier {20}(6.0,3.5)(10.5,1.5)(12.0,7.0)}
\savebox{\Vur}(72,42)[bl]{\multiput(0,0)(12,7){6}{\usebox{\wigur}}}
\savebox{\wigr}(12,0)[bl]
   {\bezier{20}(0,0)(3, 4)(6,0)
    \bezier{20}(6,0)(9,-4)(12,0)}
\savebox{\Vr}(36,0)[bl]{\multiput(0,0)(12,0){3}{\usebox{\wigr}}}
\savebox{\wigu}(0,12)[bl]
   {\bezier{20}(0,0)( 4,3)(0,6)
    \bezier{20}(0,6)(-4,9)(0,12)}
\savebox{\Vu}(0,36)[bl]{\multiput(0,0)(0,12){3}{\usebox{\wigu}}}
\savebox{\Fu}(0,48)[bl]
{ \put(0,0){\vector(0,1){27}} \put(0,24){\line(0,1){24}} }
\savebox{\Fr}(36,0)[bl]
{ \put(0,0){\vector(1,0){20}} \put(18,0){\line(1,0){18}} }
\savebox{\Fl}(36,0)[bl]
{ \put(36,0){\vector(-1,0){20}} \put(18,0){\line(-1,0){18}} }
\savebox{\Fur}(36,12)[bl]
{ \put(0,0){\vector(3,1){20}} \put(18,6){\line(3,1){18}} }
\savebox{\Fdr}(36,12)[bl]
{ \put(36,0){\vector(-3,1){21}} \put(18,6){\line(-3,1){18}} }
\savebox{\Lur}(36,12)[bl]
{ \put(0,0){\line(3,1){20}} \put(18,6){\line(3,1){18}} }
\savebox{\Ldr}(36,12)[bl]
{ \put(36,0){\line(-3,1){21}} \put(18,6){\line(-3,1){18}} }
\bma
\barr{lll}
\begin{picture}(120,70)
\put(90,33){$\gamma$}
\put(48,12){\circle*{4}}
\put(48,60){\circle*{4}}
\put(84,12){\circle*{4}}
\put(84,60){\circle*{4}}
\put(102,18){\circle*{4}}
\put(102,54){\circle*{4}}
\put(12,12){\usebox{\Fr}}
\put(12,60){\usebox{\Fl}}
\put(48,12){\usebox{\Fu}}
\put(48,12){\usebox{\Vr}}
\put(48,60){\usebox{\Vr}}
\put(84,60){\usebox{\Fur}}
\put(84,48){\usebox{\Ldr}}
\put(93,57){\vector(-3,1){3}}
\put(111,51){\vector(-3,1){3}}
\put(84,12){\usebox{\Lur}}
\put(93,15){\vector(3,1){3}}
\put(111,21){\vector(3,1){3}}
\put(84,00){\usebox{\Fdr}}
\put(102,18){\usebox{\Vu}}
\end{picture}
&\qquad&
\begin{picture}(120,60)
\put(78,29){$\gamma$}
\put(48,12){\circle*{4}}
\put(48,60){\circle*{4}}
\put(30,12){\circle*{4}}
\put(84,12){\circle*{4}}
\put(84,60){\circle*{4}}
\put(102,54){\circle*{4}}
\put(12,12){\line(1,0){36}}
\put(21,12){\vector(1,0){3}}
\put(39,12){\vector(1,0){3}}
\put(12,60){\usebox{\Fl}}
\put(48,12){\usebox{\Fu}}
\put(48,12){\usebox{\Vr}}
\put(48,60){\usebox{\Vr}}
\put(84,60){\usebox{\Fur}}
\put(84,48){\usebox{\Ldr}}
\put(93,57){\vector(-3,1){3}}
\put(111,51){\vector(-3,1){3}}
\put(84,12){\usebox{\Fur}}
\put(84,00){\usebox{\Fdr}}
\put(30,12){\usebox{\Vur}}
\end{picture}
\earr
\ema
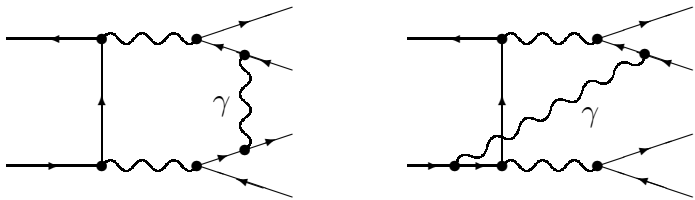
\caption{Examples of non-factorizable resonant photonic corrections}
\label{FIphodia}
\efi

\chapter{Recent Developments}

In this section we summarize the developments in the field of
radiative corrections to $\eeffff$ achieved by the European working
group on electroweak gauge bosons since the 1991 workshop \cite{EE50091}.

By a systematic expansion
of the existing complete virtual and soft-photonic $\Oa$ corrections,
a high-energy approximation has been constructed for the process
$\Pep\Pem\to\PWp\PWm$ in the limit $s, |t|, |u| \gg M_{\PW,\PZ}^2\gg
m^2_{\Pe,\Pmy,\Pta,\Pd,\Pu,\Ps,\Pc,\Pb}$,
keeping \Mt\ and \MH\ arbitrary \cite{wwhe}.
For intermediate energies ($500\GeV$ -- $2\TeV$)
the high-energy approximation has been improved by exactly taking into
account the leading low-energy universal corrections.
In the angular range $-0.9\le\cos\theta_{\Pe\PW}\le0.9$
and for energies above 500\GeV,
the approximation reproduces the complete results for all relevant
polarizations essentially within 1\%,
which roughly matches the expected experimental accuracy.

A semi-Monte-Carlo program, GENTLE \cite{Ba93}, has been developed for
$\eeffff(\gamma)$.
It includes the lowest-order resonant diagrams and the complete
initial-state \Oa~QED corrections, defined in a gauge-invariant
way through the current-splitting technique.
In addition soft-photon exponentiation and the universal $s$-channel
\Oaa~QED corrections to the initial state, known from LEP1, have been
taken into account.
The program generates distributions based on a compact analytical
formula for $d^3\si/(dk_+^2dk_-^2ds')$, with $s'=s-2\sqrt{s}E_\ga$,
and hence does not yet allow studies of angular distributions.
A generalization to angular distributions is in progress.

A Monte-Carlo event-generator, WOPPER \cite{Ma93}, has been created for
$\eeffff(n\ga)$.
Using techniques known from parton-shower algorithms,
it calculates initial-state QED corrections in the leading-log
approximation, with resummation to all orders and soft-photon
exponentiation. It is based on the resonant diagrams and provides
full spin transmission to the final state. Moreover it yields exclusive
photons with full kinematics.

\chapter{Outlook}
Finally we would like to summarize what, in our opinion, still has to
be completed to arrive at a Monte-Carlo generator that meets all
requirements.
The full lowest-order process $\eeffff$ should be calculated, including
all non-resonant contributions and also anomalous couplings. Already at
this level the leading corrections should be implemented. For the
remaining \Oa~corrections it is sufficient to evaluate the
doubly-resonant contributions. These involve electroweak as well as QCD
corrections. Of course all relevant leading
higher-order effects should be taken into account.

\end{document}